# Ultimate confinement of phonon propagation in silicon nano-crystalline structure

Takafumi Oyake[1], Lei Feng[1], Takuma Shiga[1], Masayuki Isogawa[2], Yoshiaki Nakamura[2], and Junichiro Shiomi[1,3,*]

[1]*Department of Mechanical Engineering, The University of Tokyo, 7-3-1 Hongo, Bunkyo, Tokyo 113-8656, Japan*

[2] *Graduate School of Engineering Science, Osaka University, 1-3 Machikaneyama, Toyonaka, Osaka 560-8531, Japan*

[3] *CREST, Japan Science and Technology Agency, 4-1-8, Kawaguchi, Saitama 332-0012, Japan*

[*]E-mail: shiomi@photon.t.u-tokyo.ac.jp

Temperature-dependent thermal conductivity of epitaxial silicon nano-crystalline (SiNC) structures composed of nanometer-sized grains separated by ultra-thin silicon-oxide ($SiO_2$) films (~ 0.3 nm) is measured by the time domain thermoreflectance technique in the range from 50 to 300 K. Thermal conductivity of SiNC structures with grain size of 3 nm and 5 nm is anomalously low at the entire temperature range, significantly below the values of bulk amorphous Si and $SiO_2$. Phonon gas kinetics model, with intrinsic transport properties obtained by first-principles-based anharmonic lattice dynamics and phonon transmittance across ultra-thin $SiO_2$ films obtained by atomistic Green's function, reproduces the measured thermal conductivity without any fitting parameters. The analysis reveals that mean free paths of acoustic phonons in the SiNC structures are equivalent or even below half the phonon wavelength, i.e. the minimum thermal conductivity scenario. The result demonstrates that the nanostructures with extremely small length scales and controlled interface can give rise to ultimate classical confinement of thermal phonon propagation.

PACS number(s): 63.22.-m, 65.80.-g, 63.20.K-

There is a strong need for thermally insulating dense materials in various applications such as thermoelectric energy conversion and thermal barriers. While amorphous material is known to possess low thermal conductivity due to lattice disorders, the quest in thermal science and engineering has been to achieve lower thermal conductivity based on crystal materials. Crystals have wider variability in lattice structures, and thus, when forming nanostructures, can give rise to interfaces with strong phonon reflection. Particularly for thermoelectrics[1–5], being crystal is important for mutual adoptability with high electrical conductivity and Seebeck coefficient. Over the last decides, lowering thermal conductivity by nanostructring materials with intrinsically high electronic properties has greatly enhanced thermoelectric figure of merit ($ZT$). However, since popular/common base-materials contain rare metals or toxic elements, one of the key challenges now is to realize it with abundant and ecofriendly elements. Crystal silicon (Si) is a high potential material in this sense, and is also attractive for industrial use due to compatibility with the current Si technology.

$ZT$ of Si-based thermoelectrics in general has been relatively low because of the high bulk thermal conductivity (140 $Wm^{-1}K^{-1}$ at 300 K) despite its high power factor. On the other hand, high thermal conductivity materials can benefit the most from the nanostructuring approach, and enhancement of $ZT$ has been achieved in Si nanowires[6,7], nanomeshes[8], noninclusions[4,9] and nanocrystal composites[10,11]. The extent of thermal conductivity reduction has been widely discussed in reference to Casimir limit[12], where effective scattering rate of a phonon becomes equal to its velocity divided by the nanostructure length scale (i.e. frequency of phonon encountering the interface). Therefore, reduction is larger for smaller nanostructures and denser interfaces, to which sintered nanocrystalline structures have an advantage in practice[10,13].

Physical and chemical structures at the interface naturally play an important role in the reduction. In this sense, Casimir limit is no longer an appropriate reference since it does not account for frequency dependent phonon transmission/reflection at the interface that can be strongly altered by the interfacial structures. Thermal transport across a sintered Si-Si interface has been investigated by Sakata *et al.*[14], and SiO$_x$ nano-precipitates formed at the interface were found to give rise to an order-of-magnitude controllability. Hence, there is a room to further reduce thermal conductivity by nanometer sized grains and precisely controlled interfaces, however, sintering, being a high energy process, has limitation due to the grain growth and interfacial diffusion/mixing.

An alternative is to turn to a bottom-up approach. Nakamura *et al.*[15] realized an epitaxial Si nanocrystalline (SiNC) composed of oriented Si nanoparticles covered by ultrathin $SiO_2$. The nanometer-sized grains with identical crystal orientation are separated by ultrathin $SiO_2$ (~0.3 nm or 1 monolayer thick) and connected with each other through the nanowindows (< 1 nm in diameter). The thermal conductivity of SiNC with grain size of 3 nm was measured to be about 1 $Wm^{-1}K^{-1}$, beating the amorphous limit (~2 $Wm^{-1}K^{-1}$). Now the interest is to understand the microscopic mechanism of how such low thermal conductivity is realized. In this work, we do so by measuring the temperature dependence of thermal conductivity of SiNC structures with diameters of 3 nm, 5 nm, and 40 nm in a range from 50 to 300 K, and by analyzing the data with a phonon gas kinetics model that incorporates phonon transmission/reflection at the ultrathin $SiO_2$ layer in addition to phonon-phonon scattering.

The details of fabrication method of SiNC structure has been described in Ref. 15 and Supplementary Materials[16]. Fig. 1a shows a cross sectional image of 5-nm SiNC sample, and the inset shows higher magnification images to identify the 5 nm grain size (images of SiNCs with other diameters are detailed in Ref. 15). Note that since the nanograins are connected through the nanowindows maintaining the crystal orientation, electrons can travel with coherent wavefunctions through the sample[15]. If properly doped, the structure may also manifest characteristics of granular materials[22]. However, as the purpose of the current work is to investigate solely the phonon transport, the sample was not doped and is essentially an insulator, and thus, electron contribution to thermal conductivity is negligible. The thickness of the 3-nm, 5-nm, and 40-nm SiNC samples ($h_{SiNC}$) are 20±2.2 nm, 58±2.3 nm, and 109±6.2 nm, respectively.

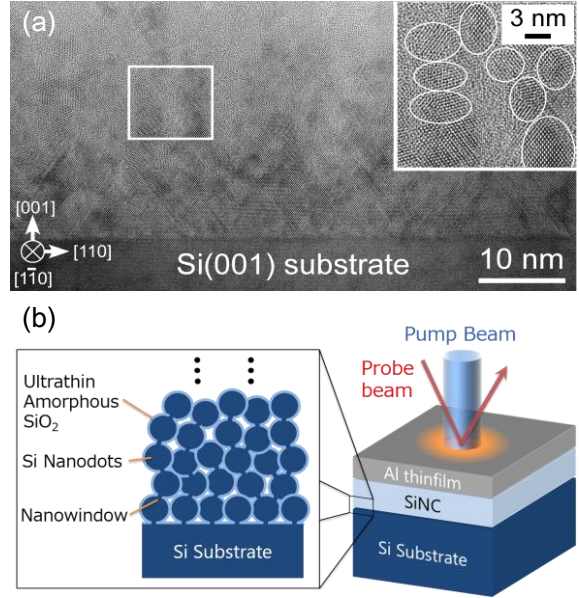

**Fig. 1** (a) Cross-sectional high resolution transmission electron microscopy image of Si nanocrystalline (SiNC) structure with a grain size of 5 nm. The inset in (a) is an enlarged image of the region marked with a rectangle. (b) Schematics of time-domain thermoreflectance (TDTR) measurements.

Thermal conductivity of SiNC thin films are measured using time-domain thermoreflectance (TDTR)[23,24]. TDTR is a well-established method that operates by pulsed laser and pump probe technique[23,24] to characterize thermal transport of thin films and interfaces. The SiNC films are coated with Al transducer film for the TDTR measurements, as shown in Fig. 1b. In our TDTR, a Ti:Sapphire laser (pulse width 140 fs, repetition rate 80 MHz ) is split into a pump pulse train (wavelength 400 nm, $1/e^2$ radius 10 μm, modulation frequency 11 MHz) and a probe pulse train (wavelength 800 nm, $1/e^2$ radius 10 μm). The pump pulse induces impulse responses at the sample surface, and the probe pulse detects the temperature change of Al transducer through thermoreflectance. Reflected probe beam consists of the in-phase voltage ($V_{in}$) and out-of-phase voltage ($V_{out}$) detected by a Si photodiode connected to a lock-in amplifier picking out the signal at the modulation frequency. The power of pump and probe beams is adjusted so that the temperature rise of sample due to steady-state heating does not exceed 2 K for all measurements. Low temperature measurements are conducted in an optical cryostat with liquid helium under vacuum below $10^{-4}$ Pa.

Thermal properties of SiNC are obtained by fitting $-V_{in}(t)/V_{out}(t)$ with a thermal model consisting of three layers: Al transducer, SiNC thin film, and Si substrate. Bulk heat capacity values are used for the Al and Si substrate. Since the SiNC structure is mostly composed of Si, bulk heat capacity of Si is used also for SiNC. The thicknesses of Al and SiNC are determined by picosecond acoustic measurement and TEM imaging, respectively. The thermal conductivity of Si substrate is measured separately on reference samples without SiNC thin film. Here, the measured thermal conductivity (resistivity) of SiNC includes the thermal boundary conductance (resistance) between SiNC and Si substrate as ultrathin $SiO_2$ between SiNC and Si substrate is also one component of SiNC structure. Note that because of the small sensitivity of TDTR signal to the thermal boundary conductance between SiNC and Si substrate (See Fig. S1 in Supplementary Material[16]), whether to include it or not does not change the values of SiNC thermal conductivity regardless of the temperature. Then, the remaining unknown parameters are thermal boundary conductance across Al and SiNC ($G_{Al/SiNC}$) and thermal conductivity of SiNC ($\kappa_{SiNC}$) (See Supplementary Materials[16] for $G_{Al/SiNC}$).

Figure 2 summarizes the obtained temperature dependence of $\kappa_{SiNC}$. The values are compared with thermal conductivities of amorphous Si and amorphous $SiO_2$. As reported previously, $\kappa_{SiNC}$ at room temperature is anomalously low (1.09 $Wm^{-1}K^{-1}$) compared with $\kappa$ of amorphous Si and amorphous $SiO_2$[15]. Furthermore, our measurements demonstrate that $\kappa_{SiNC}$ is significantly lower than those of the amorphous materials at low temperatures. Remarkably, $\kappa_{SiNC}$ with grain size of 3 nm approaches the value of minimum thermal conductivity $\kappa_{min}$. Note that the Casimir limit predicts that $\kappa_{SiNC}$ for grain size of 3 and 5 nm at the room temperature are 3.6 and 5.6 $Wm^{-1}K^{-1}$, respectively. Note here that the length scale used to calculate the Casimir limit is the grain size of each case.

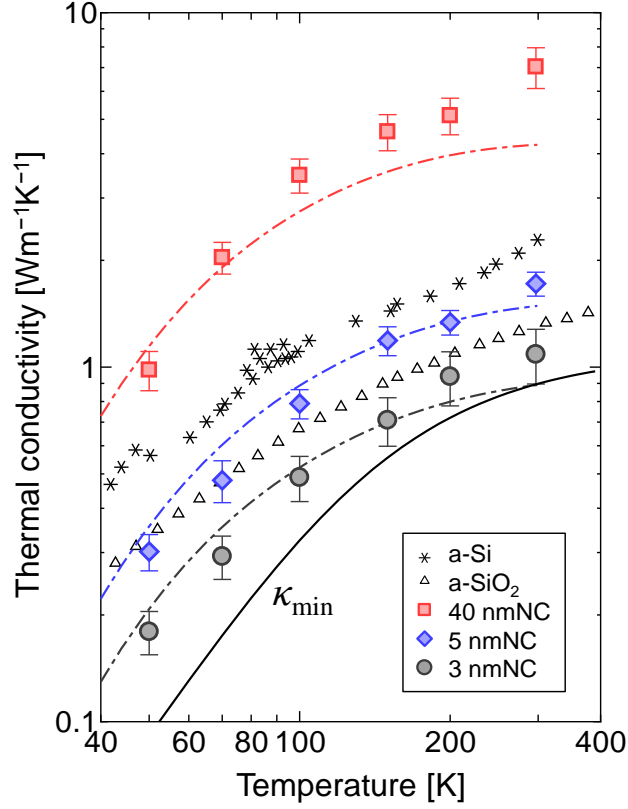

**Fig. 2** Temperature-dependent thermal conductivity of SiNC structures compared with amorphous Si (a-Si, asterisks)[44,45] and SiO$_2$ (a-SiO$_2$, triangles)[32]. Black solid line represents the minimum thermal conductivity ($\kappa_{min}$) model[32]. Dotted lines show thermal conductivity calculated by phonon gas model with inputs from anharmonic lattice dynamics and atomistic Green's function (AGF) calculations. The error bars of SiNC represent the standard deviation of measurements taken on different locations and uncertainties in Al and SiNC thicknesses and $G_{Al/Si}$.

It should be worth noting that we observe no signature of contribution from phonon interference in reducing thermal conductivity in this temperature range. Such coherence effect is expected to become larger with increasing distance between interfaces i.e. grain size in the current system, but the measured $\kappa_{SiNC}$ is lower for smaller grain size at all temperatures.

In order to gain microscopic understanding in the thermal conductivity reduction, we analyze the result in terms of kinetics of phonon gas. The solution of linearized Boltzmann transport equation gives,

$$\kappa = \frac{1}{3V}\sum_{\mathbf{k},s} C_{\mathbf{k},s} v_{\mathbf{k},s}^2 \tau_{\mathbf{k},s} \qquad (1)$$

where $V$, $s$, $\mathbf{k}$, $C$, $v$, and $\tau$ are volume, phonon polarization, wave vector, heat capacity, group velocity, and relaxation time, respectively. By adopting the Matthiessen's rule[25], $\tau$ of SiNC can be written as,

$$\tau_{\mathbf{k},s}^{-1} = \tau_{\mathbf{k},s,ph}^{-1} + \tau_{\mathbf{k},s,\text{bdy}}^{-1} \qquad (2)$$

where $\tau_{ph}$ and $\tau_{\text{bdy}}$ are relaxation times for phonon-phonon scattering and boundary scattering at SiNC interfaces, respectively. For the internal properties, $C$, $v$, and $\tau$, we use bulk single-crystal Si properties. The bulk properties were obtained by anharmonic lattice dynamics based on first-principles interatomic force constants (See Supplementary Materials[16] for details).

Since Casimir limit does not explain thermal conductivity of polycrystalline structure, we modeled $\tau_{s,\text{bdy}}$ by using the following analytical model[26],

$$\tau_{s,\text{bdy}}^{-1} = v_s \left[ (1.12 D)^{-1} + \left( \frac{\frac{3}{4} t_{\text{int}}}{1 - t_{\text{int}}} D \right)^{-1} \right] \qquad (3)$$

where $D$ and $t_{\text{int}}$ are grain size of polycrystalline structure and phonon transmittance at grain boundaries, respectively. This model with diffuse boundaries was recently validated by the Monte Carlo ray tracing simulation to be accurate even with presence of a grain size distribution with standard deviation of $0.35D$[27]. The standard deviation of grain-size distribution in the cases of 3-nm SiNC is roughly $0.25D$, and thus the Eq. (3) is applicable.

Phonon transmittance across grain boundaries in Eq. (3) is modeled as[26,28],

$$t_{int}(\omega) = \frac{1}{\gamma \omega / \omega_{max} + 1} \tag{4}$$

where $\gamma$ is a constant and $\omega_{max}$ is the maximum phonon frequency. In order to justify this model and determine the value of $\gamma$, we calculated phonon transmittance across ultrathin $SiO_2$ interfaces by the atomistic Green's function (AGF) method[29,30]. The details of AGF simulation is described in Refs. 29 and 30. In our simulation, as shown in Fig. 3a, 0.36-nm-thick or 0.72-nm-thick $SiO_2$ thin film is sandwiched between crystal-Si leads along the $z$ direction. The periodic boundary condition is applied in the directions of the cross section (the $x$ and $y$ directions) whose size is 2.17×2.17 nm$^2$. The contribution of phonons in two-dimensional Brillouin zone corresponding to the $x$ and $y$ directions is accounted by a transverse wave vector ($k_x = 10$, $k_y = 10$) grid, which was verified to ensure convergence. The interatomic interaction of Si/O atoms is modeled by Tersoff potential with parameterization by Munetoh *et al*[31]. The $\beta$-cristobalite $SiO_2$ is chosen as the initial structure of the thin film due to its smallest lattice-constant mismatch with crystal Si. The $SiO_2$ thin film was then annealed and equilibrated at 300 K, resulting in formation of an amorphous-like structure.

The simulation result shown in Fig. 3b reveals that the $SiO_2$ thin film with thickness of 0.36 nm and 0.72 nm have almost the same transmittance. $SiO_2$ thickness does not affect the thermal conductance because Si/$SiO_2$ boundary resistance dominates over the internal resistance of $SiO_2$ film when it is ultrathin (< 0.7 nm). Using this transmission function, thermal boundary conductance $G$ can be calculated in Landauer formalism with 4 probe approach[27],

$$G = \frac{\sigma}{1 - \sigma/\sigma_a} \tag{5}$$

where $\sigma$ and $\sigma_a$ are

$$\sigma = \frac{1}{2V} \sum_{\mathbf{k},s} \eta \omega \frac{\partial f}{\partial T} v_z t_{int} \tag{6}$$

$$\sigma_a = \frac{1}{2V} \sum_{\mathbf{k},s} \hbar\omega \frac{\partial f}{\partial T} v_z \qquad (7)$$

Here, $V$ is the volume, $\hbar$ is the reduced Plank constant, $f$ is Bose-Einstein distribution, and $T$ is absolute temperature. Substitution of the above obtained properties gives $G$=330 MWm$^{-2}$K$^{-1}$ independently of the SiO$_2$ thickness. In addition, solid line in Fig. 4b shows that the empirical model in Eq. (4) with $\gamma = 7.45$ produces well the AGF transmittance in the low frequency regime. Although there is some discrepancy in the high frequency regime, thermal boundary conductance calculated by Eq. (4) (345 MWm$^{-2}$K$^{-1}$) agrees with the AGF simulation result. Therefore, for the sake of simplicity, we adopt Eq. (4) with $\gamma = 7.45$ for $t_{int}$ in Eq. (3) when calculating the phonon scattering rate due to ultrathin SiO$_2$ interfaces.

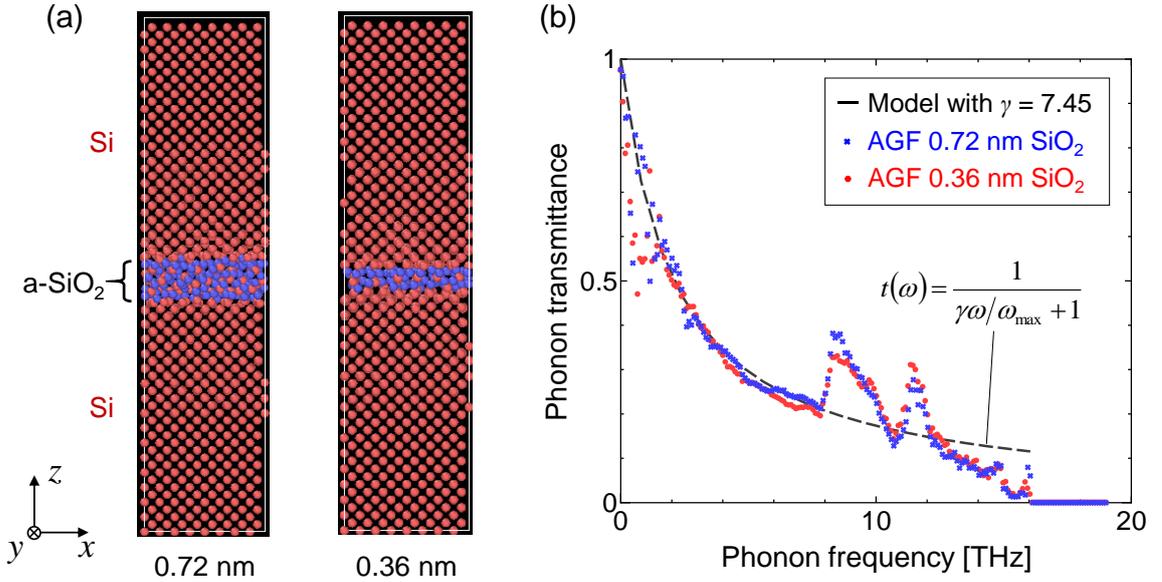

**Fig. 3** (a) Cross sectional images of amorphous SiO$_2$ (thickness of 0.72 and 0.36 nm) sandwiched by Si leads in AGF simulations. Red and blue dots represent Si and O atoms, respectively. (b) Phonon transmittance across ultrathin SiO$_2$ interface (blue x-marks: 0.72-nm SiO$_2$, red dots: 0.36-nm SiO$_2$). Dashed line represents fitting result of Eq. (3).

As shown in Fig. 2, the calculated thermal conductivity agrees well with the experimental values particularly for the cases of 3 nm and 5 nm SiNC. Note that, as described above, the calculation involves no variable parameters to be adjusted to the experiments. Furthermore, the thermal conductivity does not depend on the film thickness. This was checked by adding the boundary scattering term $4v_{k,s}/3h_{SiNC}$, and its impact was confirmed to be negligible due to the large internal thermal resistance of SiNC even for the thinnest film of 3-nm grain size ($h_{SiNC}$=20 nm). This, together with the isotropy of the grain geometry, also justifies the isotropic expression in Eq. (1). The relatively larger discrepancy in case of 40-nm SiNC can be due to the larger standard deviation of grain size distribution as the validity of Eq. (3) is expected to gradually degrade as the standard deviation exceeds $0.35D$[27]. The agreement with calculations and experiments demonstrate that the drastic reduction of thermal conductivity in SiNC structure can be described in terms of phonon scattering by ultrathin $SiO_2$ films at the SiNC interfaces.

Now that the calculation has been shown to reproduce the experiment, we investigate how mode-dependent phonon transport properties are modulated by the nanostructure. Figures 4a and 4b compare the relaxation times of phonons in SiNC structures and in single-crystal Si at 300 K and 50 K, respectively. The relaxation times in the SiNC at each temperature is drastically suppressed from the values of bulk Si due to the grain boundary scattering. A surprising feature here is that the relaxation time reaches and even becomes smaller than $\tau = \pi/\omega$, which means that mean free path of a phonon is equivalent to or smaller than half of its wavelength. Thermal conductivity calculated assuming all the phonons having such mean free path has been named the minimum thermal conductivity[32]. Important observation is minimum propagation is realized particularly for low frequency acoustic phonons with large potential contribution to thermal conductivity. On the other hand, high frequency optical phonons still have relaxation time larger than $\pi/\omega$, however, those phonons are much less dispersive than low frequency phonons, and thus, have limited contribution to the thermal conductivity. This results in the fact that thermal conductivity of 3-nm SiNC is somewhat larger but similar to the minimum thermal conductivity[32].

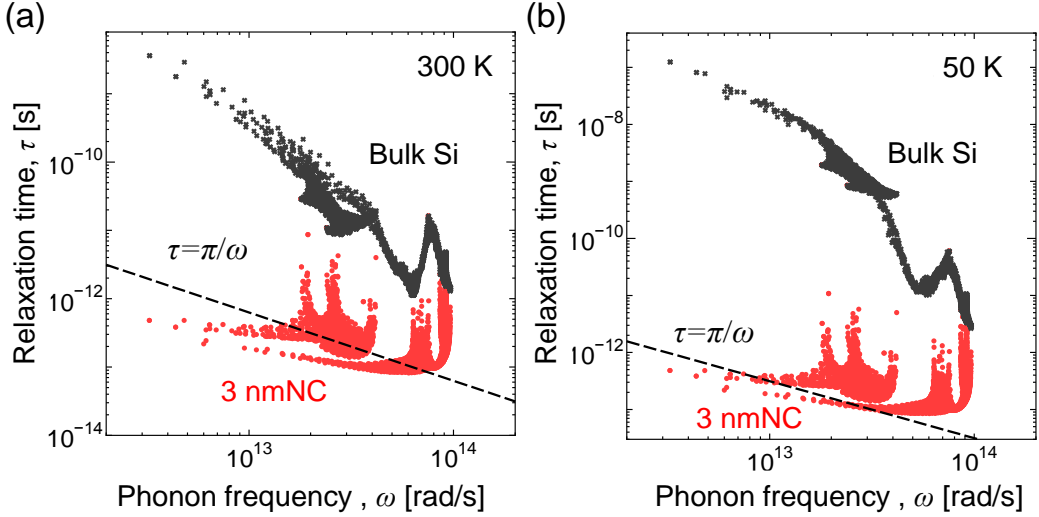

**Fig. 4** (a) Frequency-dependent phonon relaxation times of bulk Si (black x-marks) and SiNC with grain size of 3 nm (red dots) at (a) 300 K and (b) 50 K. Dashed line represents the Cahill-Pohl model for minimum thermal conductivity ($\tau = \pi/\omega$).

The indication from the mean free path smaller than half the phonon wavelength is worth discussion. This happens for some lowest frequency modes with wavelength exceeding the grain size (Fig. 4). The current kinetic model assumes all the phonons to have the states of bulk Si. There have been extensive discussions on applicability of bulk phonon properties in nanostructures, particularly for Si nanowires, in the effort to reproduce the temperature dependences of thermal conductivity measured in experiments[33] with calculations using either full[34] or bulk dispersion relations[35,36], however, the geometrical criteria for bulk phonon properties to be relevant is still unclear. The key issue here is whether phonons remain coherent across the grain, and how long-wavelength phonons can extend across the ultrathin $SiO_2$ layer. To gain insights into these issues, we have performed molecular dynamics simulations for a model system, a periodic Si(3 nm)/$SiO_2$(0.36 nm) structure (Fig. S2(a)), representing the SiNC sample, to investigate how much phonon density of states (DOS) change from that of bulk. Note that molecular dynamics include lattice anharmoncity thus incorporates the above issue of finite coherent length. As shown in Fig. S2(b), the resulting partial DOS of Si atoms is similar to that of bulk Si, indicating that the phonon states of SiNC are not largely changed from those of bulk. This should be because the $SiO_2$ layer is very thin, allowing the phonon modes to extend across the grain boundary. Although this may not be enough to prove the correctness of using bulk phonon properties, either way, the current analysis suggests that modal thermal conductivity of these lowest frequency phonons is reduced to what would be carried by the corresponding bulk-state phonons with mean free path of less than half the wavelength.

The reason why thermal conductivity becomes lower than amorphous Si lies in the low frequency modes. Although thermal energy in the amorphous materials is mainly transported by non-propagating diffusons[37], it is known that there are also low frequency propagons[32] with mean free path exceeding tens of nanometers[38-42]. Such low frequency propagation is absent in SiNC because mean free paths of acoustic phonons are ultimately reduced as described above. Possibility of reducing thermal conductivity below amorphous limit by scattering the propagons with grain boundaries has been recently analyzed by molecular dynamics simulations of the nanometer-sized polycrystalline Si nanowire[43]. The current SiNC structure is an ultimate case of such nanometer-sized polycrystalline Si as the ultra-thin $SiO_2$ layer realizes interfacial phonon reflection that is stronger than bare Si interface.

In summary, mechanisms of exceptionally low thermal conductivity of epitaxial SiNC structure was clarified by temperature-dependence measurements and rigorous phonon kinetic model. The experiments reveal that thermal conductivity of SiNC structures with grain sizes of 3 nm and 5 nm is significantly lower than amorphous Si and $SiO_2$ in the entire temperature range. By using the frequency-dependent phonon transmittance across ultrathin $SiO_2$ interface layer calculated by atomistic Green's function, thermal conductivity of SiNC can be reproduced by phonon gas kinetic model without any fitting parameters. The analysis reveals that relaxation time equivalent to or even below that of the minimum thermal conductivity scenario is realized in the 3-nm SiNC structure by extremely effective grain boundary scattering, resulting in extremely strong classical confinement of phonon propagation. This result demonstrates controllability of thermal conductivity of dense crystal materials far beyond Casimir and amorphous limits.

This work was partially supported by CREST "Scientific Innovation for Energy Harvesting Technology" (Grant No. JPMJCR16Q5) and Grant-in-Aid for Scientific Research (B) (Grant No. JP16H04274) from JSPS KAKENHI, Japan. T.O. was financially supported by Japan Society for the Promotion of Science (JSPS) Fellowship (26-9110). This work was also supported by a Grant-in-Aid for Scientific Research A (Grant No. 16H02078), a Grant-in-Aid for Exploratory Research (Grant No. 15K13276).